\documentstyle[prl,aps,twocolumn,epsf,floats]{revtex}
\input epsf
\draft
\begin{document}

\twocolumn[\hsize\textwidth\columnwidth\hsize\csname@twocolumnfalse%
\endcsname
\title{Order from disorder: Quantum spin gap in magnon spectra of LaTiO$_3$}
\author{G.~Khaliullin
}
\address{Max-Planck-Institut f\"ur Festk\"orperforschung, Heisenbergstrasse 1,
 D-70569 Stuttgart, Germany}
\date{\today}
\maketitle

\begin{abstract}
A theory of the anisotropic superexchange and low energy 
spin excitations in a Mott insulator with $t_{2g}$ orbital 
degeneracy is presented. We observe that the spin-orbit
coupling induces frustrating Ising-like anisotropy terms in the spin 
Hamiltonian, which invalidate noninteracting spin wave theory. 
The frustration of classical states is resolved by an order from disorder 
mechanism, which selects a particular direction of the staggered moment
and generates a quantum spin gap. The theory explains well
the observed magnon gaps in LaTiO$_3$. 
As a test case, a specific prediction is made
on the splitting of magnon branches at certain momentum directions.    
\end{abstract}
\draft
]

The recent renaissance in the study of transition metal oxides has 
emphasized the important role being played by the orbital  
degeneracy inherent to perovskite lattices (see for review
~\cite{TOK00,KUG82}). First of all,
the type of spin structure and the character of spin excitations crucially
depend on the orientation of occupied orbitals~\cite{GOO55,KAN59}. 
Second, the excitations in the orbital sector get coupled to the other
degrees of freedom (electronic, lattice, spin) and might therefore
strongly modify their excitation spectrum. Some examples are 
the anomalous magnon softening ~\cite{KHA00a} and incoherent
charge transport ~\cite{ISH97} due to low-energy orbital 
fluctuations in ferromagnetic manganites, and the formation 
of orbital polarons~\cite{KIL99}. 

The calculation of the spin excitation spectrum 
in systems with orbital degeneracy 
is somewhat involved even at the half filled, insulating limit.
That is due to the peculiar frustrations of superexchange interactions
~\cite{FEI97}, which lead to the infrared divergences
when a linear spin wave theory is applied. In systems with $e_g$ orbital
degeneracy the frustration is resolved by a specific, directional orbital
ordering thus breaking the cubic symmetry of spin exchange bonds~\cite{KHA97}.
The transition metal oxides with $t_{2g}$ orbitals exhibit 
different and more challenging phenomena. This occurs due to the relative 
weakness of the Jahn-Teller coupling in this case, and due to the higher
degeneracy and additional symmetry of $t_{2g}$ orbitals \cite{KHA00}. As a
result, the orbitals may form the novel, {\it coherent 
orbital-liquid\/} state stabilized by quantum effects, as observed in spin 
the $S=1/2$ Mott insulator LaTiO$_3$~\cite{KEI00}.

LaTiO$_3$ shows a spin order of G-type with magnon excitations characteristic
for a simple cubic lattice. Despite three dimensionality of magnon
spectra, the spin 
reduction is unusually large (much larger than in two dimensional cuprates), 
which has been explained in terms of fluctuating orbital 
state in this material~\cite{KEI00,KHA00}. The present paper concerns with the
origin of the small spin gap observed in LaTiO$_3$. This would not be an issue
in a conventional case with static orbital order: The latter lowers a
symmetry of the crystal and induces a spin anisotropy 
via the spin-orbit
coupling, resulting naturally in a classical gap in the magnon spectra. Orbital
order is however not observed in LaTiO$_3$, and the way how the staggered
moment chooses one out of three equivalent cubic axes as the easy one is not
that obvious. Indeed, it is shown below that the anisotropic Hamiltonian 
induced by spin-orbit coupling has a perfect cubic symmetry 
in the orbital liquid state, and a conventional spin wave 
theory gives in fact no magnon gap. 
We argue that the magnon gap and the selection of the easy magnetization 
axis in LaTiO$_3$ are nontrivial effects of quantum origin generated by 
the order from disorder mechanism.  
 
We begin with the derivation of the effective spin Hamiltonian in 
$t_{2g}$ systems. Quite generally, it consists of the 
isotropic superexchange of 
Heisenberg form, and the anisotropic spin exchange Hamiltonian. 
The latter, which is of our present interest, results from higher 
order processes involving both the spin-orbit coupling  
\begin{equation}
H_{so}= -\lambda\sum_{i}
({\bf S}_i{\bf l}_i)
\end{equation}
and the isotropic superexchange. Here the operator ${\bf l}_i$ is the
effective angular momentum ($l=1$) of $t_{2g}$ level, and ${\bf S}$ being a
spin one-half of the Ti$^{3+}$ ion. The isotropic
superexchange in orbitally degenerate systems strongly depends on the 
orbital structure. In general it can be written as: 
\begin{equation}
H_{\rm SE}^{ij}= \frac{4t^2}{U}[({\bf S}_i{\bf S}_j +\frac{1}{4})
\hat J_{ij}^{(\gamma)} ~ + ~\frac{1}{2}\hat C_{ij}^{(\gamma)}] ~,
\end{equation}
where the orbital operators $\hat J_{ij}^{(\gamma)}$ and $\hat
C_{ij}^{(\gamma)}$ depend on bond directions $\gamma(= a,b,c)$. 
In a $t_{2g}$ system like the titanates they are given by
the following expressions:
\begin{eqnarray}
\hat J_{ij}^{(\gamma)} & = &
\case{1}{2}(r_1+r_2)(P_{ij}+K_{ij})^{(\gamma)}  \nonumber \\
& & -\case{1}{3}(r_2-r_3)P_{ij}^{(\gamma)}
-\case{1}{4}(r_1-r_2)(n_i+n_j)^{(\gamma)}, \\
\hat C_{ij}^{(\gamma)} & = &
\case{1}{2}(r_1-r_2)(P_{ij}+K_{ij})^{(\gamma)}  \nonumber \\
& & +\case{1}{3}(r_2-r_3)P_{ij}^{(\gamma)}
-\case{1}{4}(r_1+r_2)(n_i+n_j)^{(\gamma)}, 
\end{eqnarray}
The coefficients 
$r_1=1/(1-3 \eta)$, $r_2=1/(1- \eta)$ and $r_3=1/(1+2 \eta)$
originate from the Hund's splitting of the excited $t_{2g}^2$ multiplet
via $\eta=J_H/U$. The operators $P_{ij}^{(\gamma)}$, $K_{ij}^{(\gamma)}$ 
and $n_i^{(\gamma)}$ can conveniently be
represented in terms of spinless fermions 
$a_i$, $b_i$, $c_i$ corresponding to $t_{2g}$ levels of $yz$, $xz$, $xy$
symmetry, respectively. 
(This notation is motivated by the fact 
that each $t_{2g}$ orbital is orthogonal 
to one of the cubic axes $a$,$b$,$c$).
Namely,
\begin{eqnarray}
P_{ij}^{(c)} & = & n_{ia}n_{ja}+n_{ib}n_{jb}, \\
K_{ij}^{(c)} & = & a_i^\dagger b_i b_j^\dagger a_j
 + b_i^\dagger a_i a_j^\dagger b_j, \nonumber \\
n_i^{(c)} & = & n_{ia} + n_{ib} \nonumber
\end{eqnarray}
for the pair along the $c$ axis.
Similar expressions are obtained for the exchange bonds along the axes $a$ 
and $b$, by replacing orbital fermions $(a, b)$ in Eqs.(5) 
by $(b, c)$ and $(c, a)$ pairs, respectively.
Since the parameter $\eta \equiv J_H/U \sim 0.16$ only~\cite{MIZ96},
we will neglect the Hund's coupling corrections. This results in a simple
expressions as obtained in~\cite{KHA00}:
\begin{eqnarray}
\hat J_{ij}^{(c)} & = & n_{ia}n_{ja}+n_{ib}n_{jb}
 + a_i^\dagger b_i b_j^\dagger a_j
 + b_i^\dagger a_i a_j^\dagger b_j, \\ 
\hat C_{ij}^{(c)} & = & -\case{1}{2}(n_{ia}+n_{ib}+n_{ja}+n_{jb}).
\end{eqnarray}
The Eq.(7) is actually unessential for our 
purpose. Summed over bonds, this term gives
the energy of the classical N\'eel state.

To complete the definitions, we express the angular momentum operator
in Eq.(1) as follows:
\begin{equation}
 l_i^z= i(a_i^\dagger b_i- b_i^\dagger a_i) , 
 l_i^x= i(b_i^\dagger c_i- c_i^\dagger b_i) , 
 l_i^y= i(a_i^\dagger c_i- c_i^\dagger a_i) , 
\end{equation}
and notice finally that the fermions must satisfy a local 
constraint, $n_{ia}+n_{ib}+n_{ic}=1$.

\begin{figure}
\epsfxsize=0.55\hsize \epsfclipon \centerline{\epsffile{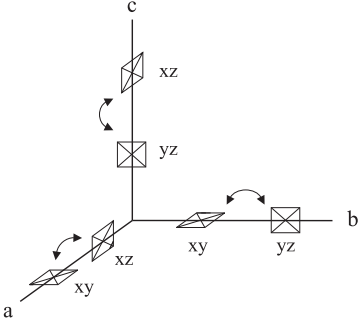}}
\vspace*{2ex}
\caption{On every bond of cubic crystal, two out of three $t_{2g}$ orbitals 
are equally involved in the superexchange and may resonate.
The same two orbitals select also a particular component of angular
momentum.}
\label{fig1}
\end{figure}
To proceed further we need to specify the orbital state. 
The two most important features of the interactions 
in the present model can be observed from Eqs.(2,6): 
{\it i})The classical N\'eel state (consider $\left\langle{\bf S}_i
{\bf S}_j\right\rangle =-1/4$) is infinitely degenerate in the orbital sector,
thus strong spin fluctuations must be present in the ground state to
provide a dynamical splitting of orbital levels. 
{\it ii})Special to $t_{2g}$ systems, every bond is represented by a
particular pair of equivalent orbitals (see Fig.1 and Eq.(6)), 
which may form dynamical orbital
singlet or triplet bonds. (In fact, the orbital operators $\hat
J_{ij}^{(\gamma)}$ may be represented in a SU(2) symmetric form~\cite{KHA00}).
This additional symmetry brings up an intrinsic quantum dynamics into 
$t_{2g}$ orbital physics, contrasting it to a classical 
behaviour of $e_g$ systems.
As described in~\cite{KHA00}, a composite spin-orbital excitation (an analogy
to a SU(4) excitation~\cite{LI98}), which consists of a spin fluctuation
accompanied by the formation of dynamical orbital bonds, best optimizes the
superexchange energy. In that orbital liquid picture cubic symmetry remains
unbroken. Angular momentum {\bf l} is however fully quenched in the ground
state, and it's fluctuation spectrum extends up to characteristic energies of
about $W_{orb}\sim 4t^2/U$. Last important remark concerns with spin-orbital 
separation: While the large energy ($\sim W_{orb}$) behaviour of the model 
Eq.(2) is governed by a coupled spin-orbital dynamics, the underlying weak
magnetic order results in a separation of low energy spin excitations
forming magnons on scale $J$. Here $J$ is a pure spin exchange coupling,
$J({\bf S}_i{\bf S}_j)$, which is obtained by averaging Eq.(6) over fast
orbital dynamics. It has been found that indeed $J\ll W_{orb}$~\cite{KHA00},
and this justifies the adiabatic approximation used below to derive an
effective spin anisotropy Hamiltonian.

By symmetry, the anisotropic pairwise interactions for spins one-half 
in a cubic crystal must have the form
\begin{equation}
H_{ani}=A(\sum_{\langle ij\rangle_a}S_i^xS_j^x~+~
\sum_{\langle ij\rangle_b}S_i^yS_j^y~+~
\sum_{\langle ij\rangle_c}S_i^zS_j^z).
\end{equation}
This interaction follows in the present model from spin-orbit corrections
to the superexchange, shown in Fig.2. Only two of them, diagrams 
a) and c) do actually contribute to the anisotropy constant $A$. Consider the
pair $\langle ij\rangle\parallel c$. The result is $H_{ani}^{ij}=AS_i^zS_j^z$
with $A=\lambda^2(2\Lambda_z^{(c)}-\Lambda_x^{(c)}-\Lambda_y^{(c)} )/4$, 
where
\begin{equation}
\Lambda_{\alpha}^{(c)}=
\langle l_i^{\alpha}\hat J_{ij}^{(c)}l_i^{\alpha}
+l_j^{\alpha}\hat J_{ij}^{(c)}l_j^{\alpha}
+l_i^{\alpha}l_j^{\alpha}\hat J_{ij}^{(c)}
+\hat J_{ij}^{(c)}l_i^{\alpha}l_j^{\alpha}\rangle_{orb}~.
\end{equation}
Here $\langle \cdot\cdot\cdot \rangle_{orb}$ implies the average over 
orbital fluctuations. The basic assumption in the above derivation is that 
these fluctuations are fast enough to enable us to integrate them out
as far as we are concerned with the low energy, magnon like dynamics of 
the spin operators. Next, 
it is important to observe that both $l_z$ (see Eq.(8)) and 
$\hat J_{ij}^{(c)}$ operate on the same orbital doublet ($a,b$). 
One therefore obtains $\Lambda_z^{(c)}>
\Lambda_x^{(c)}=\Lambda_y^{(c)}$. 
For the pairs $\langle ij\rangle\parallel a$
and $\langle ij\rangle\parallel b$, on the other hand, 
($b,c$) and ($c,a$) doublets are more active, resulting in $AS_i^xS_j^x$ and 
$AS_i^yS_j^y$ interactions, respectively. 
The physical picture is that the anisotropic 
spin interaction arises here as an 
indirect coupling via fluctuations of the angular momentum of $t_{2g}$ level.
In every direction one particular component among $l^{\alpha}$ in Eq.(8) 
is special, having more dispersive fluctuations in that direction, 
therefore the structure of Eq.(9) follows.         

Eq.(10) can be expressed in terms
of an on-site orbiton Green's functions which we evaluate using constant 
density of states within the orbiton bands of width $W_{orb}$, as
obtained in the large $N$ approximation~\cite{KHA00}. This gives
\begin{equation}
A=f(4t^2/U)(\lambda/W_{orb})^2~,
\end{equation}  
with the numerical factor $f\simeq 3$. We may notice, that $W_{orb}$ in 
this expression actually plays the role of a static splitting of levels
in a conventional, orbitally ordered case. The estimation in   
Ref.~\cite{KHA00} yields for the
Heisenberg exchange constant $J\simeq 0.16(4t^2/U)$ and the orbiton bandwidth
$W_{orb}\simeq 1.6(4t^2/U)$, thus the energy scale $4t^2/U\simeq 100$ meV 
follows using
$J_{exp}=15.5\pm 1$ meV in LaTiO$_3$~\cite{KEI00}. With $\lambda=
19$ meV~\cite{ABR70}, we then estimate $A\simeq4.2$ meV for
cubic anisotropy constant in LaTiO$_3$.

\begin{figure}
\epsfxsize=0.95\hsize \epsfclipon  \centerline{\epsffile{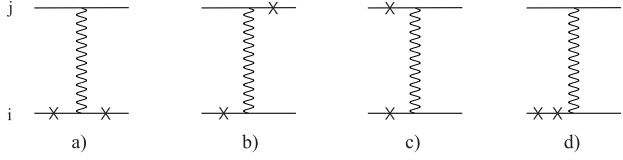}}
\vspace*{2ex}
\caption{Lowest order spin-orbit corrections to the spin exchange
Hamiltonian for a pair $\langle ij\rangle$. Wavy lines represent 
the isotropic superexchange in Eq.(2), solid lines are the orbiton 
on-site Green's functions. The spin-orbit coupling in Eq.(1) 
is denoted by crosses.}  
\label{fig2}
\end{figure}
We point out here a crucial difference between the present case and 
a canonical theory of the anisotropic superexchange interactions in orbitally
nondegenerate systems like cuprates. In $e_g$ systems, 
one has to treat both symmetric and antisymmetric 
(Dzyaloshinskii-Moriya (DM)~\cite{MOR60}) parts of the
anisotropic superexchange on equal footing because of the intimate
relationship between them~\cite{SHE92}, since both are due to the
finite deviations (tilting of octahedra etc.) from the 
ideal structure. In the present $t_{2g}$ case these interactions are, however, 
decoupled: while the DM interaction vanishes by cubic symmetry, 
a symmetric superexchange anisotropy is fully operating. The key point here 
is that the $t_{2g}$ state manifold has an intrinsic angular momentum.   

Now we consider how the interaction Eq.(9) will affect magnon excitations.
Since spin symmetry is lowered to the discrete (cubic) one, a magnon gap is
expected. A linear spin wave theory fails however to obtain it, because
Eq.(9) acquires a rotational symmetry in the limit of classical
spins. This results in an infinite degeneracy of classical states,
and an accidental pseudo Goldstone mode appears, which is however
not a symmetry property of the original quantum model Eq.(9).  
Villain's order from disorder mechanism~\cite{TSV95} comes into
play at this point, selecting a particular classical state such that the 
fluctuations about this state give the largest energy gain, 
and opening also a magnon gap. To explore this point explicitly, we calculate 
quantum corrections to the ground state energy as a function of the angle
$\theta$ between $c$ axis and the staggered moment. Assuming the latter is
perpendicular to $b$ axis, we rotate globally a spin quantization axes as
$S^z\rightarrow S^z\cos\theta-S^x\sin\theta$, 
$S^x\rightarrow S^x\cos\theta+S^z\sin\theta$,
and observe that the magnon excitation spectrum 
has an explicit $\theta$-dependence:   
\begin{eqnarray}
\omega_{\bf p}(\theta) = 
zJS\{Y_{\bf p}(X_{\bf p}+2a\eta_{zx}\sin^2\theta)\}^{1/2}~, \\ 
Y_{\bf p} = 1-\gamma_{\bf p}+a(1-c_y)~, \\ 
X_{\bf p} = 1+\gamma_{\bf p}+a(1+c_x)~.
\end{eqnarray}
Here the ratio $a=A/3J$ quantifies the deviation from the Heisenberg limit, 
and $\eta_{zx} =(c_z-c_x)/2$~, $\gamma_{\bf p} =(c_x+c_y+c_z)/3$~, 
where $c_{\alpha}=\cos p_{\alpha}$~, and $z=6$. Calculating the zero point 
magnon energy, one obtains an effective potential 
for the staggered moment, which at small $\theta$ reads as a harmonic one: 
$E(\theta) = const + K_{eff}S^2\theta^2$, 
with an effective ``spring'' constant 
$K_{eff} = \frac{1}{2}zJa\kappa \equiv A\kappa$. The parameter $\kappa$ is 
given by
\begin{equation}
\kappa = \frac{1}{S}\sum_{\bf p}\eta_{zx}(Y_{\bf p}/X_{\bf p})^{1/2}~.
\end{equation}
At small $a \ll 1$, in which case we are interested in, 
$\kappa \simeq 3aR/8S$, and $K_{eff} \simeq (R/8S)(A^2/J)$, with
the numerical factor
\begin{equation}
R = \sum_{\bf p} (1+\gamma_{\bf p}^2)(c_z^2-\gamma_{\bf p}^2)/
(1-\gamma_{\bf p}^2)^{3/2}~ \approx 0.44~.
\end{equation}

The physical meaning of the above calculation is that zero point
quantum fluctuations, generated by the interaction in Eq.(9),
are enhanced when the staggered moment stays about a symmetric position
(one of three cubic axes), and this leads to the formation of 
the energy profile of cubic symmetry.
A breaking of this discrete symmetry results then in the magnon gap, which 
should be about $\sqrt{K_{eff}/M}$ with $1/M\sim zJ$ in the harmonic 
approximation. More quantitatively, the potential 
$E(\theta)$ can be associated with an effective uniaxial anisotropy term, 
$K_{eff}\sum_{\langle ij \rangle_c} S_i^zS_j^z$,
generated in the symmetry broken phase. Therefore, one finds the magnon gap
$\Delta = zJS\sqrt{2K_{eff}/3J}$, which is a linear function of the
anisotropic interaction $A$:
\begin{equation}
\Delta = zJS\sqrt{2a\kappa} = A\sqrt{3RS}~.
\end{equation}

Now we derive this result using a different approach. Namely, we write
the equations of motion, 
$\omega S_i^{\alpha} = [S_i^{\alpha},H]$ with $H$ being the Heisenberg
exchange plus $H_{ani}$ in Eq.(9), and account for the interaction
between magnons and local fluctuations of the staggered 
moment $\delta S_j^z$, generated by $H_{ani}$, within 
the following approximation:
\begin{eqnarray}
S_i^x \delta S_j^z \Rightarrow (\chi^{(\gamma)}+\Delta^{(\gamma)})S_j^x ~, \\
\nonumber
S_i^y \delta S_j^z \Rightarrow (\chi^{(\gamma)}-\Delta^{(\gamma)})S_j^y ~.
\end{eqnarray}
Here $\gamma=a,b,c$ specifies the bond direction $\langle ij\rangle$,  
and the bond variables are defined as 
$\chi^{(\gamma)} = -\langle b_i^{\dagger}b_j\rangle$~, 
$\Delta^{(\gamma)} = -\langle b_i^{\dagger}b_j^{\dagger}\rangle$~, where
$b_i$ is the Holstein-Primakoff boson operator. This linearization
procedure leads to the magnon excitation spectrum 
\begin{equation}
\omega_{\bf p} = 
zJS\{(Y_{\bf p}+a\kappa)(X_{\bf p}+a\kappa)\}^{1/2}~, \\ 
\end{equation} 
where $\kappa = \kappa^{(a)}-\kappa^{(c)}$, 
and $\kappa^{(\gamma)} = (\chi^{(\gamma)}+\Delta^{(\gamma)})/S$.
The key observation here is that the bond variables along the 
$c$ axis differ from 
those in directions perpendicular to the staggered moment. Indeed,    
calculating bond variables $\chi^{(\gamma)}$ and $\Delta^{(\gamma)}$ 
by using the spectrum in Eq.(19), one finds $\kappa$ to be positive, 
given by the selfconsistent equation:
\begin{equation}
\kappa = \frac{1}{S}\sum_{\bf p}\eta_{zx}(Y_{\bf p}+a\kappa)^{1/2}/
(X_{\bf p}+a\kappa)^{1/2}~.
\end{equation} 
This is expected since the fluctuations originate mostly from $a$ and $b$ 
bonds in Eq.(9), once we have chosen the $c$ axis as an easy one. 
Thus $\kappa^{(a)} > \kappa^{(c)}$ follows.

Consider again the weak anisotropy case, $a\ll 1$. From equations 
(19) and (13,14) we read off the magnon gap $\Delta = zJS\sqrt{2a\kappa}$.  
Noticing also that Eq.(20) reproduces in this limit precisely 
the same parameter ${\kappa}$ defined in Eq.(15), 
we confirm finally our previous result, given by
Eq.(17), for the magnon gap. Superior to the previous case, the latter 
solution gives not only the gap value, but also the full magnon
spectrum Eq.(19), which we will discuss soon.  

One may wonder about the behaviour of Eqs.(20,19) in another extreme limit,
$a\gg 1$, where one is left with only the cubic term. Actually, the 
model defined in Eq.(9) is of certain interest by it's own~\cite{COM2}.  
For $a\rightarrow\infty$, one finds that the solution of Eq.(20) saturates
at the value
$\kappa_{\infty}\simeq (2/\pi^2S)r_S$, where $r_S=\ln(4\pi\sqrt{S})-1$.
The excitation spectrum in this limit is two-dimensional,
$\omega_{\bf p}=
2AS\{(1+\kappa_{\infty}-c_y)(1+\kappa_{\infty}+c_x)\}^{1/2}$, and it shows 
again the gap $\Delta =(4/\pi)A\sqrt{r_S S}$ of order $A$. 
The spin reduction due to the frustrations is found to be nonsingular for
finite spin values. Although it is logarithmically 
diverging in the limit of large classical spins, 
$\delta S^z \rightarrow \frac{2}{\pi^2}\ln(4\pi\sqrt{S})$,  
the ratio $\delta S^z/S$ vanishes in this limit. Therefore, we believe
that long range order in the ``cubic'' model with nearest-neighbor 
interactions is well defined for both one-half and large spins.
It should be noticed, however, that the approximation used here 
is less accurate in the ``cubic'' limit, and we anticipate substantial
incoherent features in the excitation spectrum, which is expected
to show a pseudogap of about $\Delta$ and a full 
bandwidth $\omega_{max} \sim 4AS$. 

Now we turn to the realistic case, LaTiO$_3$, and discuss the numbers.
With $A\simeq4.2$ meV estimated above, Eq.(17) gives
$\Delta\simeq3.4$ meV, which is in surprisingly good agreement
with $\Delta_{exp}=3.3\pm0.3$ meV~\cite{KEI00}.
We believe therefore that the 
quantum gap induced by the cubic anisotropy
is the dominant contribution to the magnon gap in 
LaTiO$_3$~\cite{COM3}. This can actually be tested experimentally.
The point is that the spectrum $\omega_{\bf p}$ in Eq.(19) contains 
full information of the interaction in Eq.(9) on large energy scales,
and it's peculiar structure may manifest itself in a neutron scattering 
experiment. Namely, two magnon branches of a commensurate antiferromagnet
are related to Eq.(19) as  
$\Omega_1({\bf p})=\omega_{\bf p}$ and 
$\Omega_2({\bf p})=\omega_{{\bf p}+{\bf Q}}$, where ${\bf Q}$ being 
the N\'eel vector. Because of the presence of $S^x$ and $S^y$ components in 
Eq.(9), the two branches are in fact degenerate
only in the planes with $p_x=\pm p_y$, otherwise a splitting of 
these two modes is expected at large momenta. 
The splitting is largest along $[0,\pi,p_z]$ and equivalent lines
and reaches a maximal value $2A\sim 8$ meV at the $p_z=\pi/2$ point. 
This may enable one to measure directly the value of the cubic anisotropy
constant $A$, and quantify further the orbital liquid state in a such 
puzzling material like LaTiO$_3$.

We would like to emphasize that a small value of the spin gap in LaTiO$_3$
is very difficult to explain from the traditional point of view 
(see also the discussion in~\cite{KEI00}).
Indeed, large static orbital splitting of order of 150-200 meV would be 
required in order to suppress spin-space anisotropy and hence to
explain the small magnon gaps as observed. 
However, such large splittings are very difficult to 
reconcile with the fact that neither orbital order nor any traces of 
related structural transitions have been detected. The orbital liquid concept
naturally resolves this apparent conflict: The large orbital energy scale is
generated in this picture by electronic correlations, and angular momentum
is quenched dynamically, without static symmetry breaking.

To conclude, a theory of the anisotropic superexchange of $t_{2g}$ electrons 
in titanites is presented. This interaction introduces nontrivial 
frustration in low energy spin states. We have identified a mechanism 
which selects the easy magnetic axis in cubic crystal with  
disordered orbital states. The spin gap 
in LaTiO$_3$ is most likely of quantum origin,
being thus as unique as the orbital state in this ``simple'' Mott
insulator is. An experimental test of the theory is suggested.

We would like to thank B.~Keimer for stimulating discussions.
Discussions with C.~Ulrich, R.~Zeyher and J.~van den Brink
are also acknowledged.


\begin{references}
\bibitem{COM2} Problems with this model 
have been noticed long ago: See discussion of the ``cubic'' 
model in Ref.~\cite{KUG82}, p. 253.
\bibitem{COM3} Yet a complete description of the spin anisotropy 
in LaTiO$_3$ should include also the Dzyaloshinskii-Moriya interaction
in order to explain the weak ferromagnetism of
this material. DM interaction becomes finite due to 
the small tilting of octahedra.
\end{references}
\end{document}